# A brief experience on journey through hardware developments for image processing and it's applications on Cryptography

Sangeet Saha[1], Chandrajit pal[2], Rourab paul[3], Satyabrata Maity [4], Suman Sau[5]

Dept of Computer Science & Engineering [1], A. K. Choudhury School of Information Technology[2,3,4,5]

University Of Calcutta, Kolkata, India

92, A.P.C Road, Kolkata-700009

[Sangeet.saha87, palchandrajit, rourab.paul, satyabrata.maity, sumansau] @gmail.com

## Abstract:

*The importance of embedded applications on image and video processing, communication and cryptography domain has been taking a larger space in current research era. Improvement of pictorial information for betterment of human perception like deblurring, de-noising in several fields such as satellite imaging, medical imaging etc are renewed research thrust. Specifically we would like to elaborate our experience on the significance of computer vision as one of the domains where hardware implemented algorithms perform far better than those implemented through software. So far embedded design engineers have successfully implemented their designs by means of Application Specific Integrated Circuits (ASICs) and/or Digital Signal Processors (DSP), however with the advancement of VLSI technology a very powerful hardware device namely the Field Programmable Gate Array (FPGA) combining the key advantages of ASICs and DSPs was developed which have the possibility of reprogramming making them a very attractive device for rapid prototyping. Communication of image and video data in multiple FPGA is no longer far away from the thrust of secured transmission among them, and then the relevance of cryptography is indeed unavoidable. This paper shows how the Xilinx hardware development platform as well Mathwork's Matlab can be used to develop hardware based computer vision algorithms and its corresponding crypto transmission channel between multiple FPGA platform from a system level approach, making it favourable for developing a hardware-software co-design environment.*

## Introduction

Vision processing incorporates human perception and intelligence which makes the field most interesting to the research community as it can mimic human behaviour in the computer system by means of video surveillance system, integrating more intelligence to machines such as robots, as well as in ecology, biometrics and medical applications. Interestingly, recent NASA's mission "Curiosity" on Mars, sending valuable images and information of Mars environment in a secure communication channel, transmitted images also need to processed exhaustively to find out any vital information about Mars.

Hardware designs for image and video processing is used for faster performance rather than software, to meet the requirements of the end users, keeping its market relevancy and at the same time security is another concern, so the necessity to communicate these media data securely among multiple platforms after processing to enhance human perception and satisfaction in which our focus lies.  The basic 4 steps in image processing domain are pre-processing, segmentation, feature extraction and recognition [1] and those has been keeping their strong importance in research mostly in the case of software implementation and very few implemented on hardware.

Initial pre-processing step is carried out to enhance the quality of the original image by removing noise, unbalanced brightness etc as common interfering elements followed by segmentation where images are separated from the background into various elements with properties. Next in the feature extraction stage, extraction is performed on every detected object to reduce its information to a list of parameters storing in memory. Finally in the recognition stage a set of signals are generated using this list which constitute the upper level of processing assigning a specific meaning to every detected object.

In this paper we focused on image thresholding which is mainly used in the pre-processing and segmentation stages respectively, where our implementation is performing well enough in comparison to existing work (compared below), followed by secured transmission of the image data between multiple FPGA platforms and to the best of our knowledge this design belongs to a class of advanced implementation.

Rest of the paper consists of three sections i.e. Hardware architecture and implementation design, results and observation followed by conclusion.

*A brief theory and previous work*

**Case 1: Image thresholding as a segmentation step:**

The first stage that we can think of in all stage of image processing and analysis is image binarization (i.e. to make binary image, the image should contain any two pixel values either 0 or 1 in contrast with gray images which can contains 255 pixel values for 8 bit image) which poses as one of the serious problem in applications like machine vision, pattern recognition, target tracking and image segmentation where the gray level information is required to reduce to bi-level information.

In order to extract the useful information from an image it needs to be divided into distinct components like foreground (where pixel value is '1') and background (where pixel value is  '0') objects for further analysis where most often the gray level pixels of foreground components are quite different from that of background and in this context a very crucial and significant technique available in literature known as thresholding is applied which is the process of partitioning pixels in the images into object and background classes based upon the relationship between the gray level value of a pixel and the significant parameter threshold to separate the object from the background, finding the correct value of which to separate an image into desirable foreground and background remains a very crucial step in image processing domain [2]. Because of its efficient performance and simplicity in theory, thresholding techniques have been studied extensively and a large number of thresholding methods have been published so far [4]

 A dedicated custom hardware on FPGA can process image in real time with fairly lower processing cost and power compare to software. Field Programmable Gate Arrays (FPGAs), can be used to speed up image processing applications.  An application implemented on an FPGA can be one to two orders of magnitude faster than the same application implemented in software where parallel computation of hardware should be one of the important merit of hardware platform.

In this paper we have designed and implemented an adaptive thresholding as a function of the image pixel intensities. Finding an optimal threshold value leading to an effective binarized image requires skill as the choice of the method must be done judiciously. After an initial pre-processing of the image the thresholding has been applied where the threshold value is dependent on the nature of the image which becomes a very dominant factor at the end. The thresholding procedure is straightforward after finding its optimum value in general given by

Let $\mathbf{a} \in \mathbf{R}^n$ be the source image and $[h, k]$ be a given threshold range. The thresholded image $\mathbf{b}\ \{0, 1\}^x$ is given by

$$b(x) = 1 \ \text{if}\ h \leq a(x) \leq k$$
$$0 \quad \text{otherwise}$$

Where $\mathbf{x}$ is of form $\mathbf{x} = (x_1, x_2, \&, x_n)$, where for each $i = 1, 2, \&, n$, $x_i$ denotes a real number called the $i^{th}$ *coordinate* of $\mathbf{x}$. The most common point sets occurring in image processing are discrete subsets of $n$-dimensional Euclidean space $\mathbf{R}^n$ with $n = 1, 2,$ or $3$ together with the discrete topology. However, other topologies such as the *von Neumann topology* and the *product topology* are also commonly used topologies in computer vision [3] Otsu's thresholding computation involves many iterative complex arithmetic operations such as multiplications and divisions, which does not lend well to a high-speed and low cost implementation.[5] which is where our design is very simple with respect to both the aspects. Also according to ref no [5] Otsu's method is only used for gray-scale image segmentation, so the RGB data has to be sent into RGB2YCbCr module in order to get the image intensity data as luma component before it is sent to the main processing module whereas our model treats each individual colour channels i.e R G B as their greyscale equivalences without converting into RGB2YCbCr, in other words to its luma component Y, saving a lot of time and circuit processing complexity and loss of data as Y is the additive component of 30% of the red value, 59% of the green value, and 11% of the blue value.

In paper [6] a clustering-based method, namely, weighted artificial neural network is proposed to calculate the threshold however this approach is applicable to a specific domain and proposing a general neural network to solve all kind of problems seems to be problematic whereas our approach shows good results for different types of image like aerial, degraded documents, texture, normal colour images etc and the main disadvantage of the cluster approach is that an inappropriate choice of the number of clusters may yield poor result as the quality of the final solution depends largely on the initial set of clusters, and may, in practice, be much poorer than the global optimum. According to paper optical flow computation [7] the difference between two consecutive gray level averages (of two consecutive images), indicates the appropriate threshold, where the gray level average is indicated by

$$a = \sum_{i}^{m} \sum_{j}^{n} \frac{p(i,j)}{m \times n} \qquad \text{-----------------------------(1)}$$

Where $m \times n$ denote the grid dimension, i and j are the pixel coordinates (they are not consecutive pixels, i, j = 1k, 2k, ...k ≥ 1, if k = 1 all the image is averaged). But however if there are no changes between two consecutive images, i.e., the brightness is constant and there is no movement, then the grey level average must be the same for each image and under this situation the difference between the two average gray level is zero so as the threshold which is undesirable and this situation is likely to occur very often and also we get no information regarding the resource usage which is very important for optimising a particular design whereas in our design there is no possibility of the threshold getting zero unless and until the image behaves so and all other information have been provided.

In [8] authors had designed and implemented their optimised threshold architecture used for medical imaging applications but there was a drawback i.e. the selection of the threshold value is static which is independent of the nature of the pixel intensity and that is dynamic in our implemented design. Today's sophisticated medical imaging applications pressurises the demand of dynamic thresholding for useful information extracting as a result of segmentation. Secondly paper [8] shows only behavioural simulation and left the hardware implementation part as a future scope whereas we are successfully overcome both the above issues. The hardware implementation of the binary implementation procedure is shown in the section 3 and section 4 contains the results and observations.

*Case 2: Secured transmission of the image data between multiple FPGA platforms*

Image applications are used in internet, multimedia systems, medical, and telemedicine, military for the purpose of communication whereas this communication is not secured at all and could fall in the prey of any attacker who could attack on the transmitted images thus secretes and sensitiveness of the images can be disclosed to the unauthorized person.

Thus in this paper we described a model of Hardware architecture that could be the backbone of the main hardware model of secured image transmission.

The model contains two FPGA platforms namely the Xilinx Spartan 3E and Virtex 5 leading to a heterogeneous communication between them, and RS-232 cable for the medium of transmission. As a first step, from the host PC a pixel value of the digital image is read and it is being sent to the serial port of the PC in the Mathworks Matlab software platform. Next form the serial port of PC the image data is transmitted to the DCE port of first FPGA board which actually performing as an encryption engine by executing stream cipher encrypting algorithm named RC4 [9] .The encypted image data is then sent to the second FPGA platform which actually acts as the decryption engine by the means of execution of decryption part of RC4. The decrypted image data and the original image is viewed on a TFT monitor acting as a display unit. To accomplish this purpose we had to customize a TFT interface controller from the board to the display unit which is another merit of our paper. For the sake of verification we set up the TFT interface at the first board for observing the encrypted image as shown in fig 6 in section 4.
.

There are few papers [10, 11] available which gave the some way out to deal with secured image transmission but the basic model described here is simple, efficient and easy to implement.
Here lies the outline of the experiment being carried out.

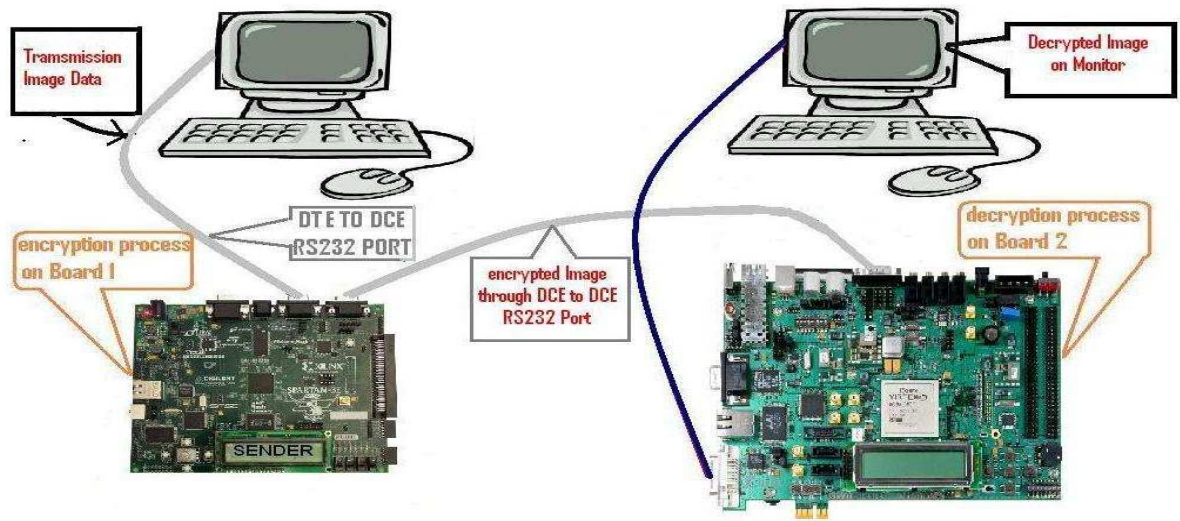

Fig 1: Secured image transmission on multiple FPGA platform

## 3. Hardware architecture and implementation deisgn:

Fig. 1 shows the structure of multiple FPGA platform communication of encrypted image data.

Fig. 2 and fig 3 shows the detailed hardware architectures of the image filtering(in two units, 1 and 2) and thresholding units respectively.
The image filtering equation with a particular kernel is as shown in equation 2.

$$f[x,y] * g[x,y] = \sum_{n_1=-\infty}^{\infty} \sum_{n_2=-\infty}^{\infty} f[n_1,n_2] \cdot g[x-n_1, y-n_2] \quad\quad\quad (2)$$

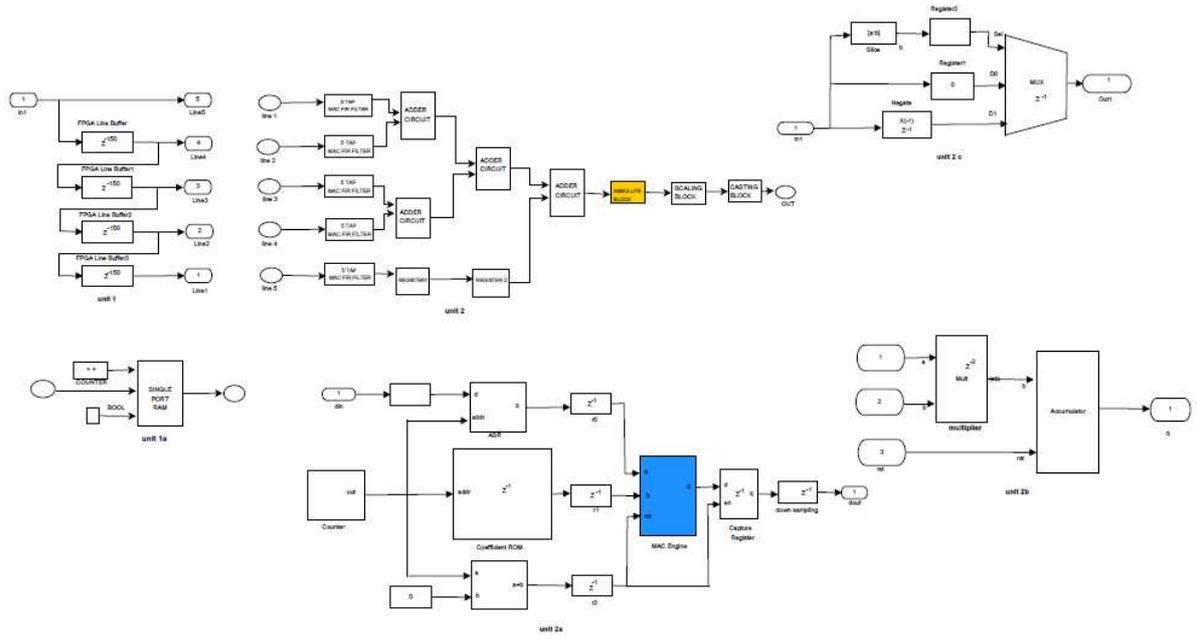

Fig 2: Filtering hardware architecture

The image filtering hardware consists of 5 buffer lines each one of which is logically selected based on the size of the filter kernel (5x5), as shown in unit 1 of Fig. 2. The buffer line consists of a single port RAM, as shown in the unit (1a) of Fig. 2; the counter in it is incremented to write the current pixel data and to read it subsequently. The output of each of five buffers of unit-1 goes to respective inputs of unit-2, each of five parallel sub-circuits of unit-2 consists of five MAC FIR engines; one such unit is elaborately shown in unit-2a depicting the ASR (Addressable Shift Register) block capable to address inputs and to incur delay at different rates. Five outputs of five MAC engines are sequentially added to get the result, whose absolute value is computed and the data is narrowed to 8-bits. The blue colored block is elaborated in unit-2b as the MAC engine. The yellow box is elaborated in unit-2c, which calculates the absolute value before multiplying with the scaling factor. In the fig 3 subsystem contains the hardware architecture shown in fig 3.

*For Image thresholding as a segmentation step:*

We have successfully tested our design on coloured as well as grayscale images First of all in case of a coloured image it is broken down into three separate matrix structure channels namely Red, Green and Blue with which the picture is composed of , for processing individually. Then each picture matrix is processed in hardware for smoothing purpose taking a 5X5 smoothing kernel over the entire picture matrix. The smoothed image is then used for thresholding calculation. The threshold for three different RGB matrices are calculated separately for binarization. The three different binarised image is cascaded following the rule of matrix concatenation. The $i_{th}$, $j_{th}$ location of each RGB stream is added up i.e ($R_{ij}+G_{ij}+B_{ij}$) followed by the expression satisfying a conditional statement the final output image is reconstructed.

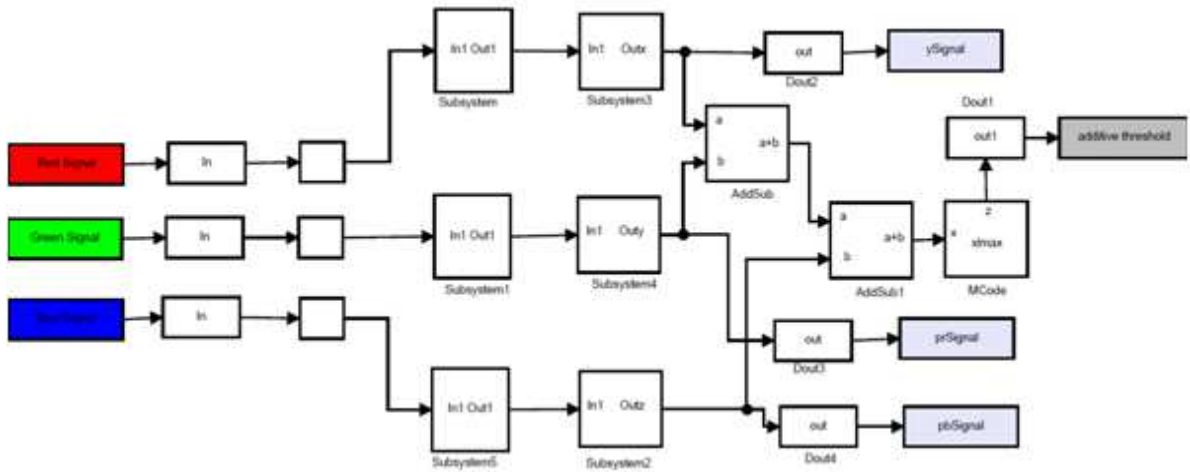

Fig 3: hardware architecture for image thresholding.

**Algorithm flow:**

1) Divide the colour image into its constituent Red, Green and blue component matrices.
   For each separate channel i.e Red, Green and Blue perform the following
2) Calculate the histogram of individual matrices.
3) Repeat for **1 to n**   // where n is the total no of different pixel intensity present//
      Multiply and accumulate a type of pixel intensity with its number of occurance.
    end
4) Repeat step 3 for red, green and blue matrices.
5) Divide step 3 with the total no of pixels present in a particular matrix to calculate the threshold for it.
6) Construct the binary image for each **R,G,B.**
7) Add the similar pixel positions for each RGB.$\sum (R_{ij}+G_{ij}+B_{ij})$ **i=1 to m and j= 1 to n**, where **m** and **n** are the total no of rows and columns...
8) If summation is less than 1 set
      Output pixel <= '0';
   Else
      Output pixel <= '1';

## 4. Results and Observation

*For image thresholding:*

Figure above is a snapshot from the WaveScope tool(ISIM simulator) which provides a powerful and easy-to-use waveform viewer for analyzing and debugging System Generator designs where the time-changing values of any wires in the design after the conclusion of the simulation have been observed. The signals may be formatted in a logic or analog format and may be viewed in binary, hex, or decimal radices. It was obtained from execution of the code. It consists of two parts viz: the left

portion showing the parameter list and the right one having the results corresponding to each of the parameters on the left-hand-side. All these values are in decimal. The RGB value is represented in decimal terms.

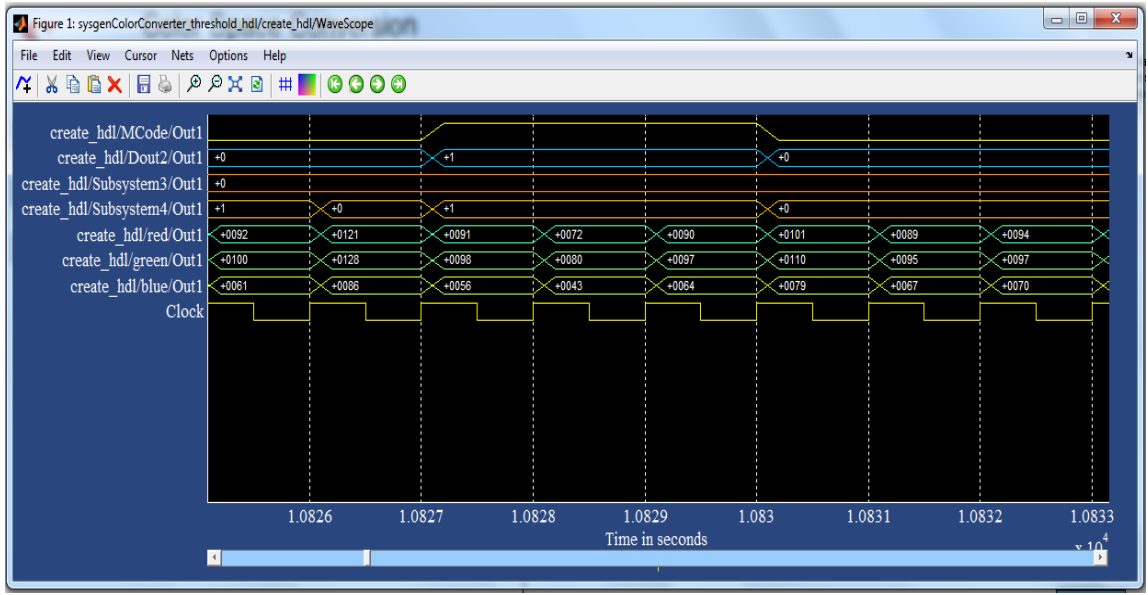

Fig 4: ISIM viewer

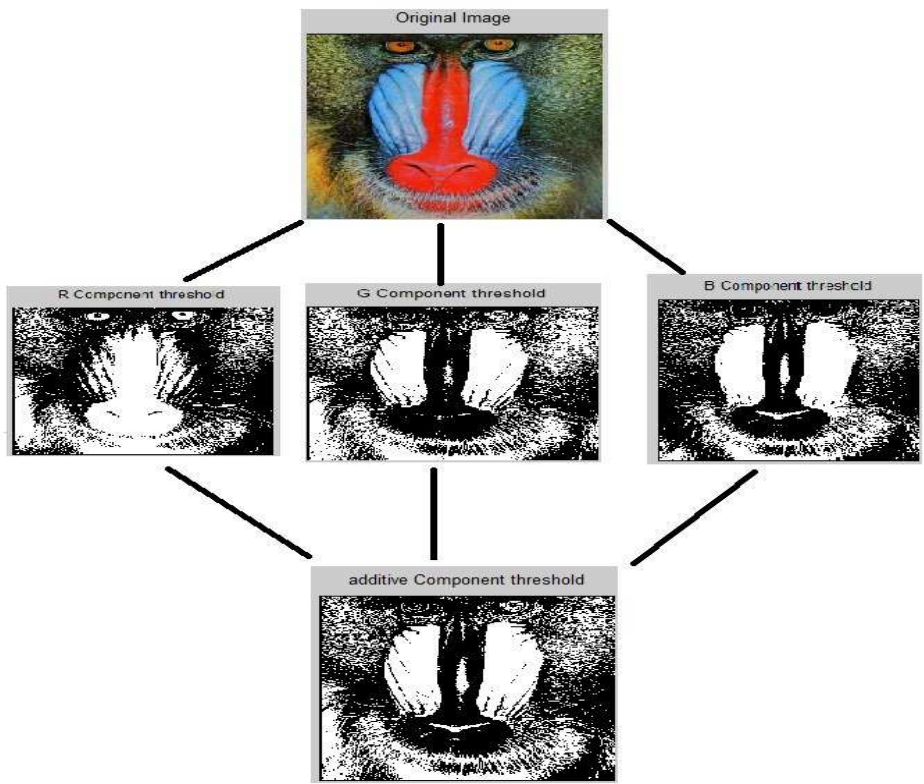

Fig 5 : image binarisation technique

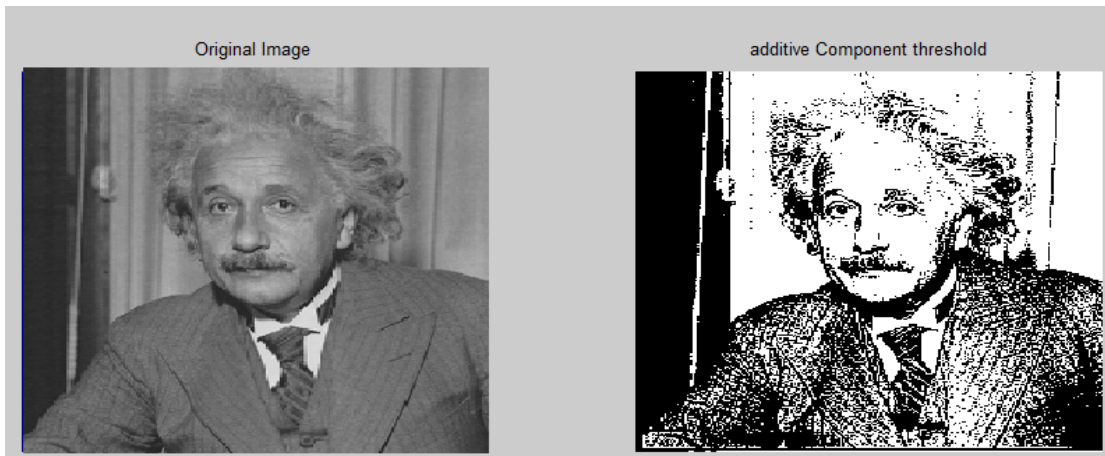

Fig 5.a

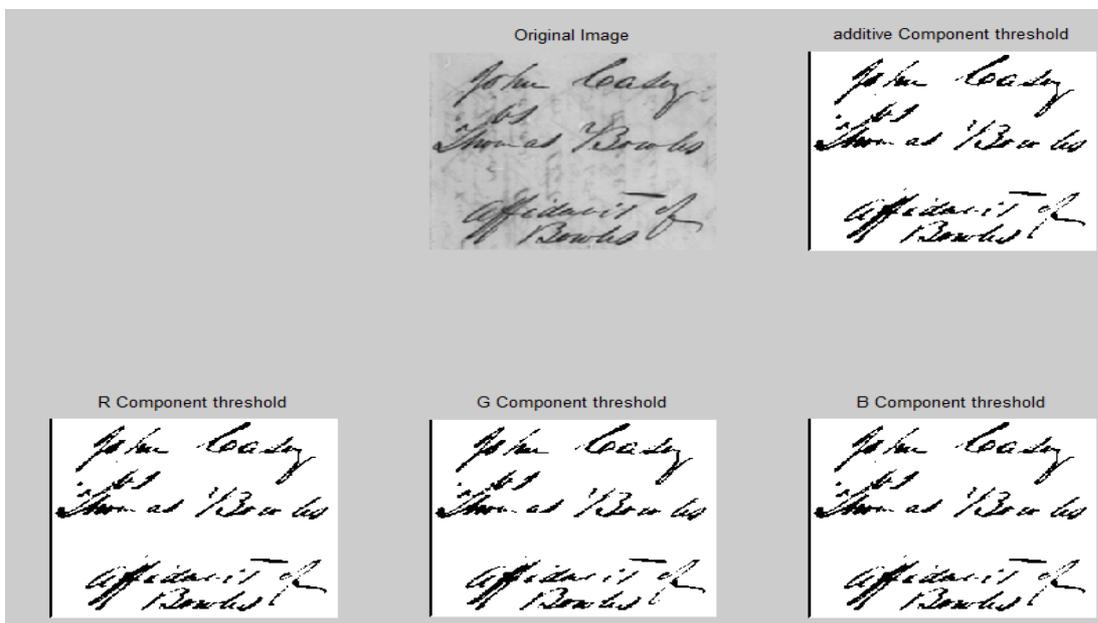

Fig 5.b: Showing improvement in the degraded documents

As shown in fig 5, a particular coloured image is broken down into 3 sub parts namely its R, G, B components and thresholding operation is carried out on the individual components which are finally added up to construct the final binarised image and if the original image is a grayscale one then only one section of the channel component is used rest remaining same.

*For secured image transmission on multiple FPGA:*

The following images are original pictures that were taken during the experiments were carrying out.

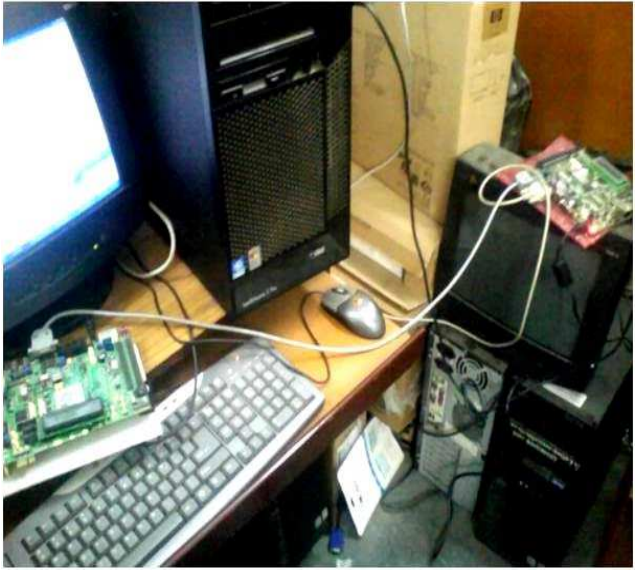
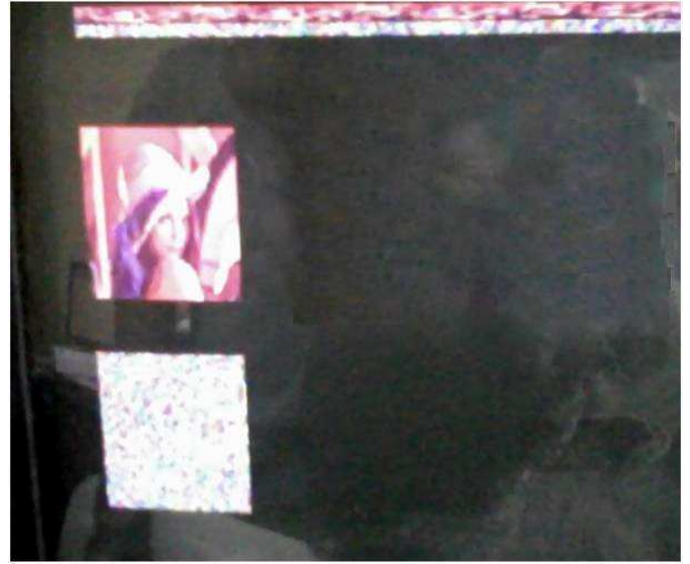

**Image -1**  **Image -2**

Fig 6: Output images for secured image transmission

Here the image-1 depicts the original view where two heterogeneous FPGA board were shown and in image-2 the original image and corresponding encrypted image has been shown.

## 5. Conclusion

Here throughout this paper we briefly discussed the work that has been carried out on image processing domain particularly by emphasizing its implementation on hardware device and also transmission of image data through a secured way. This paper will encourage the further initiatives to be taken for implementation of work in such domain. In the paper the time complexity of the whole thresholding method and limitation of image data to be stored in the FPGA board are few limitation that has to be taken care, we stared working on basic filtering mechanism, digital image water marking and some other issues related to image security and in near future we could propose some innovative idea related to this.

**Acknowledgement**

We are really grateful to our research supervisor Dr. Amlan Chakrabarti, Reader , A.K.Choudhury School of IT and Dr.Ranjan Ghosh, Associate Professor, Institute Of Radio Physics and Electronics, University of Calcutta as well as Department of Science And Technology (DST, Govt of India) for all kinds of support and encouragement to carry out this research work.